\documentclass[aps,prl,twocolumn]{revtex4-1}  
\usepackage{amsmath,amssymb}  
\usepackage{graphicx}         
\usepackage{booktabs}
\usepackage{natbib}
\usepackage[colorlinks=true, linkcolor=blue, citecolor=blue, urlcolor=blue]{hyperref} 
\usepackage{tabularx}
\usepackage{bm}
\usepackage{orcidlink}
\usepackage{braket}
\usepackage{subfig}
\usepackage{ulem}
\usepackage{ragged2e} 

\def\HIM{Helmholtz Institute Mainz, 55099 Mainz, Germany}
\def\GSI{GSI Helmholtzzentrum für Schwerionenforschung GmbH, 64291 Darmstadt, Germany}
\def\JGU{Johannes Gutenberg University, Mainz 55128, Germany}

\def\FMUV{Faculty of Mathematics, University of Vienna, Oskar-Morgenstern-Platz 1, 1090 Vienna, Austria}
\def\GPGUV{Gravitational Physics Group, University of Vienna, W\"{a}hringer Stra{\ss}e 17, 1090 Vienna, Austria}

\def\USyd{School of Physics, The University of Sydney, Sydney, New South Wales 2006, Australia}
\def\UC{Department of Physics, University of California at Berkeley, Berkeley, California 94720-7300, USA}

\begin{document}

\title{Searching for Exotic Interactions between Antimatter}

\author{Lei Cong$^{1,2,3}$\orcidlink{0000-0003-0002-1840}, Filip Ficek$^{4,5,*}$\orcidlink{https://orcid.org/0000-0001-5885-7064}, Pavel Fadeev$^{3}$\orcidlink{https://orcid.org/0000-0002-4328-6614}, Yevgeny V.~Stadnik$^{6}$\orcidlink{0000-0002-3544-5160}, 
and Dmitry Budker$^{1,2,3,7,+}$\orcidlink{0000-0002-7356-4814}}
\address{
$^{1}$\HIM\\
$^{2}$\GSI\\
$^{3}$\JGU\\
$^{4}$\FMUV\\
$^{5}$\GPGUV\\
$^{6}$\USyd\\
$^{7}$\UC\\
* filip.ficek@univie.ac.at; $^+$ budker@uni-mainz.de
 }

\begin{abstract}
We show that atomic antimatter spectroscopy can be used to search for new bosons that carry spin-dependent exotic forces between antifermions. 
A comparison of a recent precise measurement of the hyperfine splitting of the $1$S and $2$S electronic levels of antihydrogen and bound-state quantum electrodynamics theory yields the first tests of positron-antiproton exotic interactions, constraining the dimensionless coupling strengths $g_pg_p$, $g_Vg_V$ and $g_Ag_A$, corresponding to the exchange of a pseudoscalar (axionlike), vector, or axial-vector boson, respectively. 
We also discuss 
new tests of CPT invariance
with exotic 
spin-dependent and spin-independent
interactions involving antimatter. 

\end{abstract}

\maketitle



\begin{figure}[!htbp]
\includegraphics[width=0.48\textwidth]{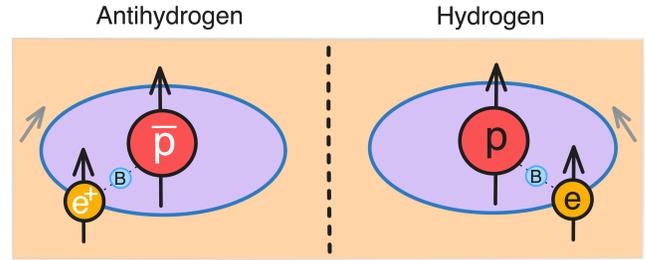}
\caption{\justifying Spin-dependent exotic interactions mediated by a new boson (B) in antihydrogen and hydrogen. 
}
\label{fig:schematic}
\end{figure}

Spin-dependent exotic interactions \cite{cong_spin-dependent_2024} are hypothetical interactions mediated by Beyond-Standard-Model (BSM) bosons, which depend on the intrinsic spins of interacting fermions.
Such interactions have been extensively studied in systems composed of ordinary stable matter, including electrons, neutrons, and protons. 
For example, exotic interactions between 
(matter) fermion pairs
such as $e$-$p$ have been studied \cite{karshenboim_precision_2010,hunter_using_2013,hunter_using_2014}. 
Much less attention has been given to exotic interactions involving matter-antimatter fermion pairs $e$-$e^+$, $e$-$\mu^+$, $e$-$\overline{p}$ \cite{karshenboim_precision_2010, ficek_constraints_2018}.
Here, 
as illustrated in Fig.\,\ref{fig:schematic}, 
we present the first study of spin-dependent exotic interactions between purely antimatter species—specifically, positron-antiproton ($e^+$-$\overline{p}$) interactions—through atomic spectroscopy of antihydrogen.

The existence of 
antimatter particles
was confirmed with the discovery of the positron (or anti-electron) by Anderson in 1932 \cite{anderson_positive_1933}. 
The antiproton was discovered later in 1955 \cite{chamberlain_observation_1955}. 
The antihydrogen atom \cite{charlton_antihydrogen_2020,khabarova_antihydrogen_2023}, consisting of one positron and one antiproton, was first artificially produced in 1995 with the help of accelerators at CERN \cite{baur_production_1996}, a result later confirmed at Fermilab in 1996 \cite{blanford_observation_1998}. 
The specialized Antiproton Decelerator (AD) facility at CERN provides unique opportunities to test fundamental physics. For example, the Antihydrogen Laser Physics Apparatus (ALPHA) experiment investigates CPT symmetry through precision spectroscopy of antihydrogen \cite{amole_experimental_2014, ahmadi_observation_2017, smorra_parts-per-billion_2017}. 
The ALPHA-g experiment paves the way for testing the Weak Equivalence Principle (WEP) by studying the gravitational behavior of antihydrogen \cite{anderson_observation_2023}. 
Moreover, it has also been pointed out that any anomalous results observed in antimatter experiments may not be due to modifications of the fundamental principles of our existing theories, but rather to new BSM interactions, so-called “fifth forces” \cite{charlton_antihydrogen_2020}.


Spin-dependent exotic interactions, i.e., 
interactions of a new boson with fermion spins, 
also known as a spin-dependent ``fifth force", have attracted increasing interest recently \cite{safronova_search_2018, cong_spin-dependent_2024}.
Theoretical studies have explored the exotic potentials arising from the exchange of a spin-0 boson \cite{moody_new_1984} and a spin-1 boson \cite{dobrescu_spin-dependent_2006,fadeev_revisiting_2019}. 
One example of the former is the axion 
\cite{weinberg_new_1978,wilczek_problem_1978,kim_weak-interaction_1979,zhitnitsky_possible_1980,shifman_can_1980,dine_simple_1981}, which is a leading candidate for dark matter \cite{preskill_cosmology_1983,abbott_cosmological_1983,dine_not-so-harmless_1983,jackson_kimball_search_2023}, a solution to the strong CP problem \cite{peccei_constraints_1977,peccei_mathrmcp_1977} (in the case of the QCD axion), and other phenomena \cite{sakharov_violation_1967}. 
Besides the canonical QCD axion, more generic spin-0 axion-like particles (ALPs) are also possible \cite{jaeckel_low-energy_2010,arvanitaki_string_2010}. 
Numerous groups actively seek axions of cosmological (dark matter), astrophysical (solar), and laboratory (virtual) origin using a range of different methods; see, for example, Refs.\,\cite{graham_experimental_2015,madmax_working_group_dielectric_2017,ouellet_first_2019,choi_recent_2021,ringwald_review_2024}. 
Spin-0 bosons can possess both scalar ($s$) and pseudoscalar ($p$) type interactions with fermions, allowing for three qualitatively different combinations of vertex types: $s$-$s$, $p$-$p$, and $p$-$s$. 
While the $s$-$s$ interaction gives a spin-independent exotic potential 
(usually denoted by $V_{ss}$)
at leading order, the other two are spin-dependent at leading order and are denoted by ${V}_{pp}$ and ${V}_{ps}$. 
As for new spin-1 bosons, possible candidates include the $Z^\prime$ boson \cite{langacker_physics_2009} and the paraphoton $\gamma'$ \cite{okun_limits_1982,holdom_two_1986,appelquist_nonexotic_2003,dobrescu_massless_2005}, 
and there are six types of interactions, depending on the Vector(V)/Axial-vector(A) and Tensor(T)/Pseudotensor($\tilde{T}$) vertex combinations, namely ${V}_{VV}$, ${V}_{AA}$,  ${V}_{AV}$, ${V}_{TT}$, $V_{\tilde{T}\tilde{T}}$ and ${V}_{\tilde{T}T}$. 


In this work, we study three of these spin-dependent exotic potentials, namely ${V}_{pp}$, ${V}_{VV}$ and ${V}_{AA}$, based on antihydrogen spectroscopy. Previous study have focused on exotic spin-independent ``fifth-force'' interactions described by a Yukawa-type potential \cite{charlton_antihydrogen_2020}, yielding constraints on long-range (Earth-scale) interactions between matter and (anti)hydrogen. In contrast, we probe atomic-scale, primarily spin-dependent interactions between antimatter particles.
Beyond this, we test CPT invariance by examining the effects of both spin-dependent and spin-independent exotic interactions in analogous transitions in antihydrogen and hydrogen. This approach is free from Standard Model theoretical uncertainties, providing a precise test of fundamental symmetries.

Atomic spectroscopy is a powerful approach in the search for spin-dependent exotic interactions. 
It is complementary to electric dipole moment searches \cite{stadnik_improved_2018}, parity violation studies \cite{antypas_isotopic_2019}, comagnetometer experiments \cite{kim_experimental_2018}, and 
spin-polarised
torsion-pendulum experiments \cite{heckel_preferred-frame_2008}, which collectively span investigations from subatomic to astrophysical domains. 
For an overview of these techniques and their applications, see Ref.\,\cite{cong_spin-dependent_2024}.

Spectroscopic experiments 
can be
highly accurate, which translates into high sensitivity to exotic interactions.
At present, this is possible only for a limited number of simple systems \cite{karshenboim_hyperfine_2011,karshenboim_constraints_2010,karshenboim_precision_2010,ficek_constraints_2017,ficek_constraints_2018,fadeev_pseudovector_2022} that consist of only a few particles, such as hydrogen \cite{karshenboim_hyperfine_2011,karshenboim_constraints_2010,karshenboim_precision_2010,fadeev_pseudovector_2022,cong_improved_2024} and ideally leptonic atoms such as
positronium \cite{frugiuele_current_2019} and muonium \cite{fadeev_pseudovector_2022}, 
which lack a usual ``hadronic'' nucleus and have much smaller theoretical uncertainties related to hadronic physics. 
A more complicated exotic atom is antiprotonic helium. 
Ficek \textit{et al.}
\cite{ficek_constraints_2018} compared the experimental measurements of transition energies between two states within the hyperfine structure (hfs) manifold of the antiprotonic helium $(n,l)=(37,35)$ state \cite{pask_antiproton_2009} with quantum electrodynamics (QED) predictions \cite{korobov_fine_2001}. 
This led to constraints on several types of exotic interactions between an electron and an antiproton.
Theoretical predictions and experimental results agree 
well on the hyperfine structure 
of antihydrogen. 
We use these predictions and results to put bounds on 
exotic spin-dependent interactions between antimatter at distances on the order of a Bohr radius or longer. 
The measured $1$S 
\cite{ahmadi_observation_2017-1} and $2$S \cite{baker_precision_2025} hfs, 
along with the $1$S-$2$S $d$-$d$ transition \cite{rasmussen_aspects_2017,ahmadi_characterization_2018} 
(see Fig.\,\ref{fig:level-structure} for a schematic 
illustration of the relevant energy levels),
provide an exceptional opportunity to compare with QED predictions and search for new physics. 

\begin{table*}[htbp]
\caption{\justifying Comparison of theoretical and experimental values to calculate $\Delta E$. 
Here $\Delta_{\text{hfs}}(1\textrm{S})$ is the $1$S hfs interval. 
$D_{21}$ is the $1$S and $2$S hyperfine interval difference, defined as $D_{21} =  8\Delta_{\text{hfs}}(2\textrm{S}) -\Delta_{\text{hfs}}(1\textrm{S})$. 
For antihydrogen, the experimental value 
of $D_{21}$
is $0.4\pm4\,\mathrm{MHz}$, which we calculate using $\Delta_{\text{hfs}}(1\textrm{S}) = 1420.4(5)\,\mathrm{MHz}$ \cite{ahmadi_observation_2017-1} and $\Delta_{\text{hfs}}(2\textrm{S}) = 177.6(5)\,\mathrm{MHz}$ \cite{baker_precision_2025}. 
The $d-d$ transition refers to the transition between the $1$S and $2$S hyperfine-split sublevels $d$ in the presence of an applied magnetic field of strength $B = 1.03285(63)\,\textrm{T}$ \cite{ahmadi_characterization_2018}. 
}\label{table1}

\begin{tabular}{lccc}
\hline
\hline
Parameter & $\Delta_{\text{hfs}}(1\textrm{S})$ &  $D_{21}$ \\
\hline
Theory (H) & 
 1420.452\,MHz \cite{karshenboim_hyperfine_2002}
& 48.9541(23)\,kHz \cite{yerokhin_electron_2008} \\
Expt ($\bar{\textrm{H}}$) & 1420.4(5) MHz \cite{ahmadi_observation_2017-1} & 0.4$\pm4$\,MHz \cite{ahmadi_observation_2017-1,baker_precision_2025} \\
\hline
$\mu$ & 0.1\,MHz & -0.35\,MHz\\
$\sigma$ & 0.5\,MHz & 4.0\,MHz\\
\hline
$\Delta E$ (95\%) &
1.0\,MHz  & 
7.9\,MHz\\
\hline
\hline
\end{tabular}
\hspace{1cm}        
\begin{tabular}{lcc}
\hline
\hline
Parameter & $d$-$d$ [at 1.03285(63)\,T]\\
\hline
Expt (H) &  2,466,061,103,080.3(6)\,kHz \cite{parthey_improved_2011,ahmadi_characterization_2018} \\
Expt ($\bar{\textrm{H}}$) &  2,466,061,103,079.4(5.4)\,kHz \cite{ahmadi_characterization_2018}  \\
\hline
$\mu$ & 0.9\,kHz\\
$\sigma$ &5.4\,kHz\\
\hline
$\Delta E$ (95\%) & 
10.8\,kHz\\
\hline
\hline
\end{tabular}
\end{table*}

Table\,\ref{table1} summarizes the latest theoretical and experimental results as well as the difference  between 
them ($\mu$),
and the combined uncertainty ($\sigma$). 
We use these to derive $\Delta E$, representing 
a measure of
the maximum 
possible
deviation \cite{ficek_constraints_2018,fadeev_pseudovector_2022}. 
The calculations are based on the following equations: $\mu = \text{Theory} - \text{Expt}$\, 
(or the experimental difference $\textrm{H}-\bar{\textrm{H}}$ in the case of the parameter $d$-$d$),
$\sigma = \sqrt{{\sigma_\textrm{th}^2} + {\sigma_\textrm{expt}^2}}$\,, $\Delta E$ is derived from the integral equation given by:
\begin{equation} I = \int_{-\Delta E}^{\Delta E} \frac{1}{\sqrt{2\pi}\sigma} e^{-\frac{(x-\mu)^2}{2\sigma^2}} dx = 0.95 \, , 
\end{equation}
where a $95\%$ confidence level \cite{cong_spin-dependent_2024} is chosen to give constraints for exotic interactions between a positron and an antiproton in this work. 
 
We use the 1S hfs and $D_{21}$ results to obtain constraints on the coupling coefficients of pseudoscalar/pseudoscalar, vector/vector, and axial-vector/axial-vector interactions, denoted as $g_p g_p$, $g_V g_V$, and $g_A g_A$, respectively. 
The hfs interval refers to the energy difference between the two hyperfine sublevels in antihydrogen, arising from the interaction between the magnetic moments of the 
antiproton
nucleus and the positron.
The hyperfine splitting differences 
$D_{21} = 8\Delta_{\text{hfs}}(2\textrm{S}) - \Delta_{\text{hfs}}(1\textrm{S})$ 
have been used to probe exotic forces in hydrogen, deuterium, and the helium-3 ion \cite{karshenboim_precision_2010,karshenboim_hyperfine_2011,fadeev_pseudovector_2022,cong_improved_2024}. 
Unlike conventional tests based on the 1s hfs alone, 
which are limited by nuclear structure uncertainties, 
$D_{21}$ is largely free from leading-order nuclear effects due to the 
8:1
scaling of electron density at the nucleus. 
For further details, see Refs.\,\cite{karshenboim_hyperfine_2011,series_spectrum_1988,karshenboim_hyperfine_2002,karshenboim_precision_2005}.
In addition, we use the hydrogen/antihydrogen difference in the $d-d$ transition to check for a possible matter-antimatter discrepancy and to constrain exotic 
interaction differences $g^e g^p - g^{{e}^+} g^{\bar{p}}$ for the axial-vector/axial-vector, pseudoscalar/pseudoscalar, vector/vector and scalar/scalar interaction combinations.

The specific spin-dependent exotic potentials studied in this work are $V_2|_{AA}$, $V_3|_{AA}$, $V_3|_{pp}$ and $(V_2+V_3)|_{VV}$, as well as the spin-independent terms $V_1|_{VV}$ and $V_1|_{ss}$. 
Based on the theoretical framework presented in Ref.\,\cite{cong_spin-dependent_2024}, these potentials take the following form: 
\begin{equation}
\label{gaga_V2}
\begin{aligned}
V_{2}|_{AA} = -g_A^{e^+}g_A^{\overline p} \frac{\hbar c}{4\pi}\left(\boldsymbol{\sigma}_{e^+}\cdot\boldsymbol{\sigma}_{\overline p} \right)\frac{1}{r}\,e^{-M c r / \hbar } \, , 
\end{aligned}
\end{equation} 
\begin{equation}
\label{gaga_V3}
\begin{aligned}
&V_{3}|_{AA}
= - g_A^{e^+}g_A^{\overline p} 
\frac{\hbar^3}{M^2 c}  
\left[ \boldsymbol{\sigma}_{e^+} \cdot \boldsymbol{\sigma }_{\overline p}  \left( \frac{1}{r^3} + \frac{Mc}{r^2\hbar} + \frac{4 \pi}{3} \delta(\boldsymbol{r}) \right)  \right. \\
&\left. - \left( \boldsymbol{\sigma}_{e^+} \cdot \hat{\boldsymbol{r}} \right) \left( \boldsymbol{\sigma }_{\overline p}  \cdot \hat{\boldsymbol{r}} \right)  \left( \frac{3}{r^3} + \frac{3Mc}{ r^2\hbar} + \frac{M^2c^2}{ r \hbar^2} \right)  \right] \frac{e^{-M cr / \hbar }}{4 \pi}\,,
\end{aligned}
\end{equation}
where \(\hbar\) is the reduced Planck constant, \(c\) is the speed of light in vacuum, \(\boldsymbol{\sigma}_{e^+}\) and \(\boldsymbol{\sigma}_{\overline p} \) are vectors of Pauli matrices representing the spins \(\boldsymbol{s}_i=\hbar \boldsymbol{\sigma}_i/2\) of the two fermions \(e^+\) and \(\overline p\), 
\(m_{e^+}\) and \(m_{\overline p}\) 
are the corresponding fermion masses, 
$M$ is the new boson mass, which is inversely proportional to the force range $\lambda$, $M=\frac{\hbar}{c\lambda}$, and \(r\) is the distance between \(e^+\) and \(\overline p\).

The potential $V_3$ also represents a dipole-dipole interaction generated by the exchange of a pseudoscalar axion or ALPs between fermions: 
\begin{equation}\label{gpgp_V3}
\begin{aligned}
&V_{3}|_{pp}=\\
&-g_p^{e^+}g_p^{\overline p}\frac{\hbar^3}{16\pi c}\frac{1}{m_{e^+} m_{\overline p}}\left[\boldsymbol{\sigma}_{e^+}\cdot\boldsymbol{\sigma}_{\overline p}   \left(\frac{1}{r^3}+\frac{M c}{ r^2 \hbar}+\frac{4\pi}{3}\delta(\boldsymbol{r})\right)\right.\\
&\left.-(\boldsymbol{\sigma}_{e^+}\cdot \hat{\boldsymbol{r}})(\boldsymbol{\sigma}_{\overline p} \cdot \hat{\boldsymbol{r}})\left(\frac{3}{r^3}+\frac{3Mc}{ r^2 \hbar}+\frac{M^2 c^2 }{ r \hbar^2}\right)\right]e^{-M c r / \hbar} \, . 
\end{aligned}
\end{equation}

The $(V_2+V_3)|_{VV}$ potential we study here has the form: 
\begin{equation}
\label{gVgV_V23}
\begin{aligned}
&(V_2+V_3)|_{VV} = 
g_V^{e^+}g_V^{\overline p}\frac{\hbar^3}{16\pi c m_{e^+} m_{\overline p}}\\
&\left[ \boldsymbol{\sigma}_{e^+} \cdot \boldsymbol{\sigma }_{\overline p}   \left( \frac{1}{r^3} + \frac{M c}{ r^2 \hbar} + \frac{M^2 c^2 }{ r \hbar^2} - \frac{8 \pi}{3} \delta(\boldsymbol{r}) \right) \right. \\
&\left.-  \left( \boldsymbol{\sigma}_{e^+} \cdot \hat{\boldsymbol{r}} \right) \left( \boldsymbol{\sigma }_{\overline p}  \cdot \hat{\boldsymbol{r}} \right)  \left( \frac{3}{r^3} + \frac{3M c}{ r^2 \hbar } + \frac{M^2 c^2 }{ r \hbar^2} \right)  \right] e^{-M c r / \hbar } \, , 
\end{aligned}
\end{equation}
while the spin-independent exotic potentials take the forms:
\begin{equation}
\label{gVgV_V1}
V_1|_{VV} = + g^{e^+}_V g^{\overline p}_V \frac{\hbar c}{4\pi}\frac{e^{-M c r / \hbar}}{r} \, , 
\end{equation}
\begin{equation}
\label{gsgs_V1}
V_1|_{ss} = - g^{e^+}_s g^{\overline p}_s \frac{\hbar c}{4\pi}\frac{e^{-M c r / \hbar }}{r} \, , 
\end{equation}
for $V_1|_{VV}$ and $V_1|_{ss}$, respectively.

\textbf{1. Constraints on exotic interactions via 1S hfs and $D_{21}$}

The exchange of a new boson via these exotic interaction potentials 
can affect
atomic spectroscopy 
measurements. 
To estimate the contribution of the energy shift 
at a magnetic-field strength of $|\boldsymbol{B}_0| = 0\,\textrm{T}$,
we use the positron wave functions of the 1S state of antihydrogen: 
\begin{equation}\label{wavefunctionmain}
\Psi_{1\textrm{S}_{a,b,c,d}}(r) = \psi_{1\textrm{S}}(r) \otimes \chi_{a,b,c,d}\,,
\end{equation}
where $\psi_{1\textrm{S}}(r) = \frac{1}{\sqrt{\pi a_0^3}} e^{-r / a_0}$ is the spatial part of the 
ground-state
wave function 
described by the quantum numbers $(n,l,m)=(1,0,0)$,
with $a_0$ being the Bohr radius; 
The term $\chi_{a,b,c,d}$ represents the spin part, corresponding to $\ket{F=0}$ and $\ket{F = 1, m_F=-1}$, $\ket{F = 1, m_F=0}$, and $\ket{F = 1, m_F= +1}$, respectively, as introduced in Appendix A.


The energy shift caused by $V_i$ for the 1S hfs 
interval
is: 
\begin{equation}\label{DeltaE1}
\begin{aligned}
\Delta E^\textrm{Exotic}_{1S,\textrm{hfs}} = \bra{\Psi_{1\textrm{S}_d}} V_i \ket{\Psi_{1\textrm{S}_d}} -\bra{\Psi_{1\textrm{S}_a}} V_i \ket{\Psi_{1\textrm{S}_a}} + \\
\bra{\Psi_{1\textrm{S}_b}} V_i \ket{\Psi_{1\textrm{S}_b}} -\bra{\Psi_{1\textrm{S}_c}} V_i \ket{\Psi_{1\textrm{S}_c}}\,,
\end{aligned}
\end{equation}
where the $1\textrm{S}_{b,c,d}$ states are degenerate 
in the absence of a magnetic field,
meaning they are identical in this context. 

\begin{figure*}[!htbp]
\includegraphics[width=0.325\textwidth]{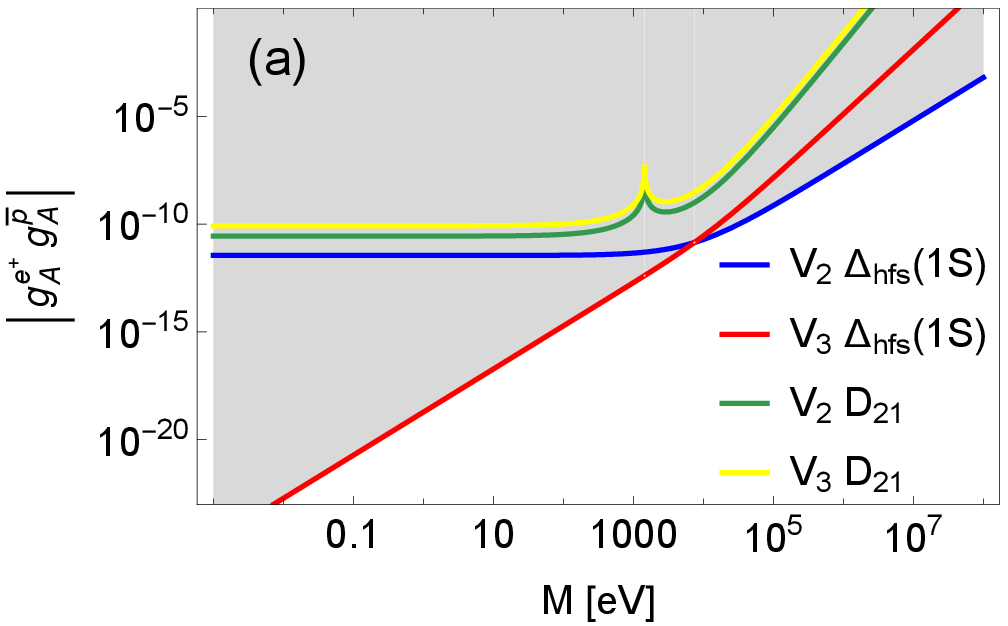}
\includegraphics[width=0.325\textwidth]{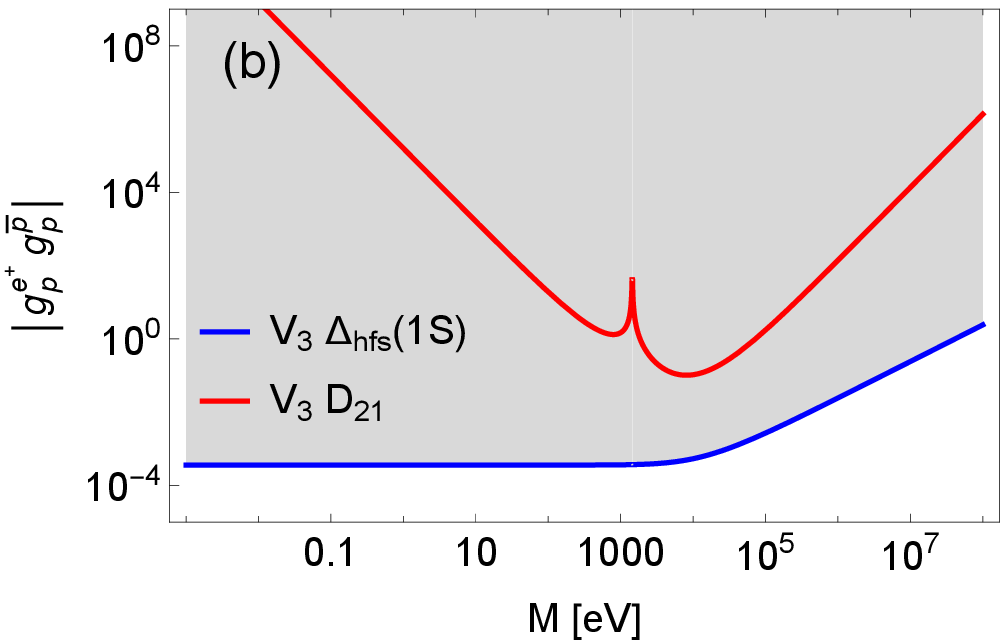}
\includegraphics[width=0.325\textwidth]{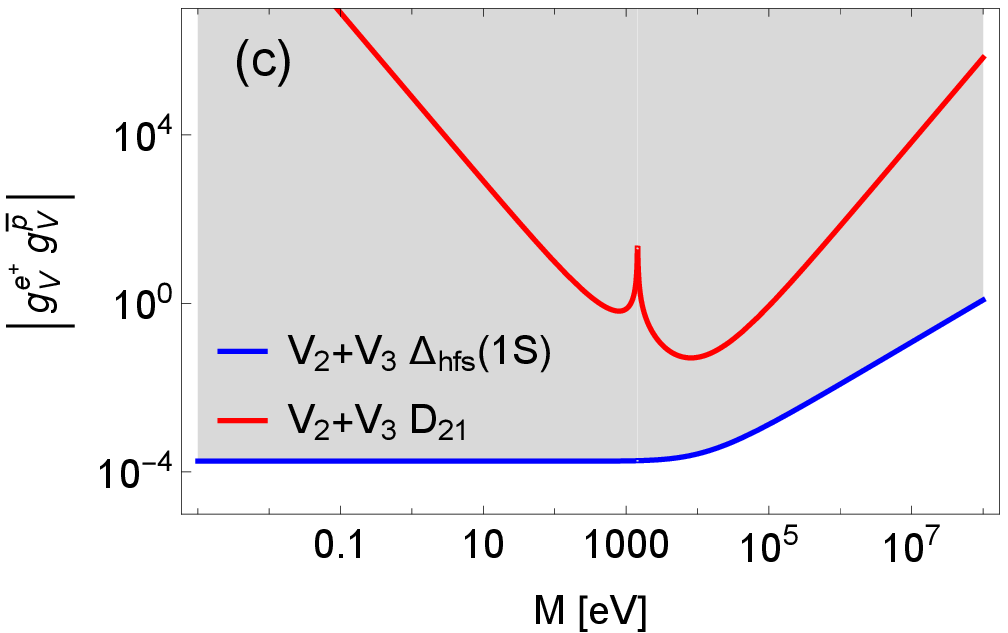}
\caption{\justifying 
Constraints on the coupling constant products 
$g^{e^+} g^{\bar{p}}$ for the (a) axial-vector/axial-vector, (b) pseudoscalar/pseudoscalar and (c) vector/vector interactions
as a function of the new boson mass $M$ (at 95\% confidence level). 
}
\label{fig:Constraints}
\end{figure*}

The ground-state hyperfine transition in antihydrogen has been measured in \cite{ahmadi_observation_2017-1}. 
The experiment measures 
the doubly-differential quantity
$(E_d - E_a) - (E_c - E_b)$ at $|\boldsymbol{B}_0| \sim 1$\,T \cite{ahmadi_observation_2017-1}. 
However, since the magnetic field correction for such an energy level difference is zero, it is considered the experimental result for the 1S hfs interval (at 0\,T).
An important detail is that the (null) correction for the finite magnetic field in this chosen combination of transitions is free from any theoretical uncertainty \cite{budker_atomic_2010}.
Using the 1S hfs transition, we obtain a maximum deviation of $|\Delta E| < 1.0$ MHz, as shown in Tab.\,\ref{table1}.

Using Eq.\,\eqref{DeltaE1} and the bound on $|\Delta E|$ (1.0\,MHz), we obtain constraints on $g_i^{e^+}g_i^{\overline p}$ as a function of the new boson mass $M$; see also the Appendix B.
The obtained bounds apply to the potentials given in Eqs.\,(\ref{gaga_V2})--(\ref{gVgV_V23}). 
The results are presented in Fig.\,\ref{fig:Constraints}\,(a-c). 
Note that while we choose to provide constraints at 0\,T, the same approach can also be used to study the $\sim$1\,T case. This can be done by evaluating the rhs of Eq.\,\eqref{DeltaE1} using the eigenstates at $\sim$1\,T. Then $\Delta E^\textrm{Exotic}_{1S,\textrm{hfs}}$ is limited by the theoretical-experimental differences at $\sim$1\,T. The theoretical result for $(E_d - E_a) - (E_c - E_b)$ remains the same as the 1S hfs interval, since the magnetic field correction is zero. 
In addition to the $1$S hfs, the specific difference $D_{21}$ \cite{karshenboim_hyperfine_2011,fadeev_pseudovector_2022,cong_improved_2024} is relatively insensitive to uncertainties in the determination of magnetic moments and other fundamental constants, making it potentially advantageous for theoretical interpretation.
Similarly, under the condition $0$\,T, using 
\begin{equation}\label{DeltaED21}
\begin{aligned}
\Delta E^\textrm{Exotic}_{D_{21}} &= 8(\bra{\Psi_{2\textrm{S}_d}} V_i \ket{\Psi_{2\textrm{S}_d}} -\bra{\Psi_{2\textrm{S}_a}} V_i \ket{\Psi_{2\textrm{S}_a}})-\\
&(\bra{\Psi_{1\textrm{S}_d}} V_i \ket{\Psi_{1\textrm{S}_d}} -\bra{\Psi_{1\textrm{S}_a}} V_i \ket{\Psi_{1\textrm{S}_a}})\,,
\end{aligned}
\end{equation}
and the bounds on $|\Delta E |$ 
(7.9\,MHz, as shown in Tab.\,\ref{table1}),
we can obtain constraints for $g_i^{e^+}g_i^{\overline p}$.
Note that $\ket {\Psi_{2\textrm{S}_d}}$ has the same form as $\ket {\Psi_{1\textrm{S}_d}}$ in Eq.\,\eqref{wavefunctionmain}, except the spatial part takes the form $\psi_{2\textrm{S}}(r) = \frac{1}{\sqrt{32 \pi a_0^3}} \left(2 - \frac{r}{a_0}\right) e^{-r / (2a_0)}$. 

Our results in Fig.\,\ref{fig:Constraints}\,(a) show that for $V_2$, the constraints from $D_{21}$ do not exceed those from $1$S hfs, while in the case of hydrogen, for $V_2$, the constraints from $D_{21}$ and the $1$S hfs are complementary, each setting the best limits for different force ranges \cite{karshenboim_precision_2010}. 
This indicates that the experimental precision for hyperfine structures in antihydrogen has not yet reached a sufficient level.
However, the $V_3$ terms provide us with strong constraints in the low-mass range. 
This is because $V_3|_{AA}$ contains a factor of $1/M^2$, as seen in Eq.\,\eqref{gaga_V3} and the theoretical works \cite{cong_spin-dependent_2024,fadeev_revisiting_2019,fadeev_pseudovector_2022,karshenboim_hyperfine_2011}. 
As shown in Fig.\,\ref{gaga_V3}\,(b,\,c), the $1$S hfs provides constraints better than the $10^{-3}$ level for $g_p g_p$ and $g_V g_V$ for new boson masses $M<10^4$\,eV. 
Additionally, existing constraints have been obtained from muonium, hydrogen, deuterium, tritium, $^3$He$^+$ ion and positronium \cite{stadnik_searching_2023,karshenboim_precision_2010,fadeev_pseudovector_2022}. 
Note that all of these systems involve different combinations of fermions; however, antihydrogen is, so far, the only purely antimatter-based atom that has been used in such searches.

\textbf{2. Testing CPT with the $d-d$ transition}

\begin{figure*}[!htbp]
\includegraphics[width=0.325\textwidth]{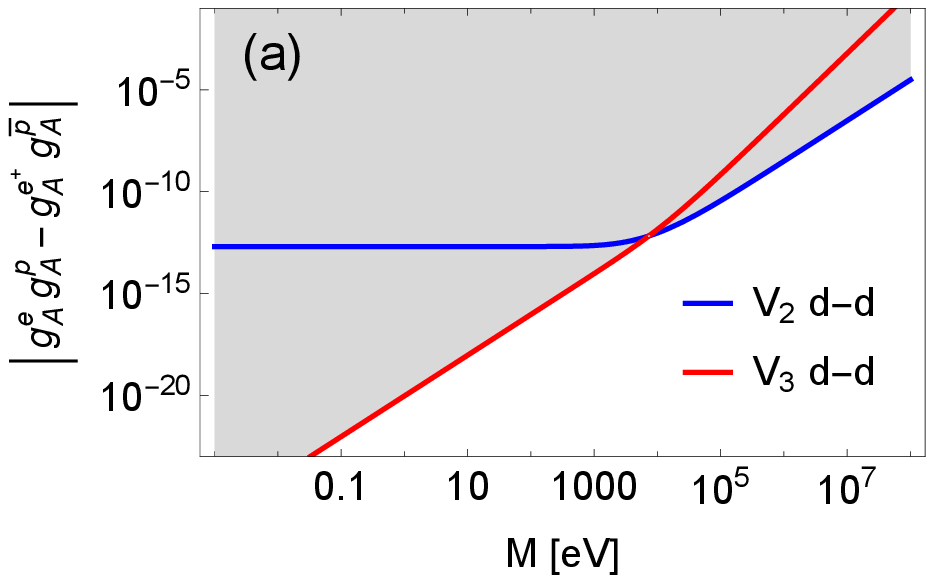}
\includegraphics[width=0.325\textwidth]{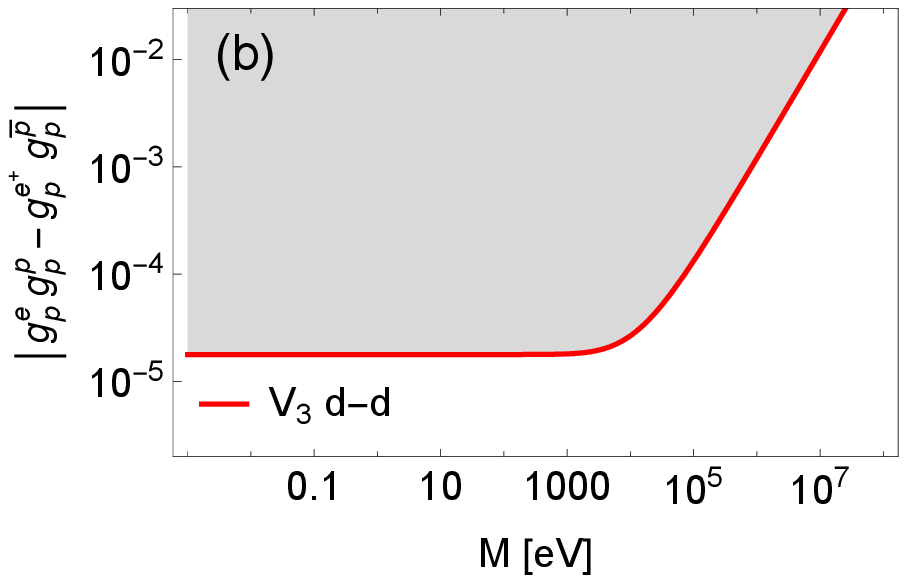}
\includegraphics[width=0.325\textwidth]{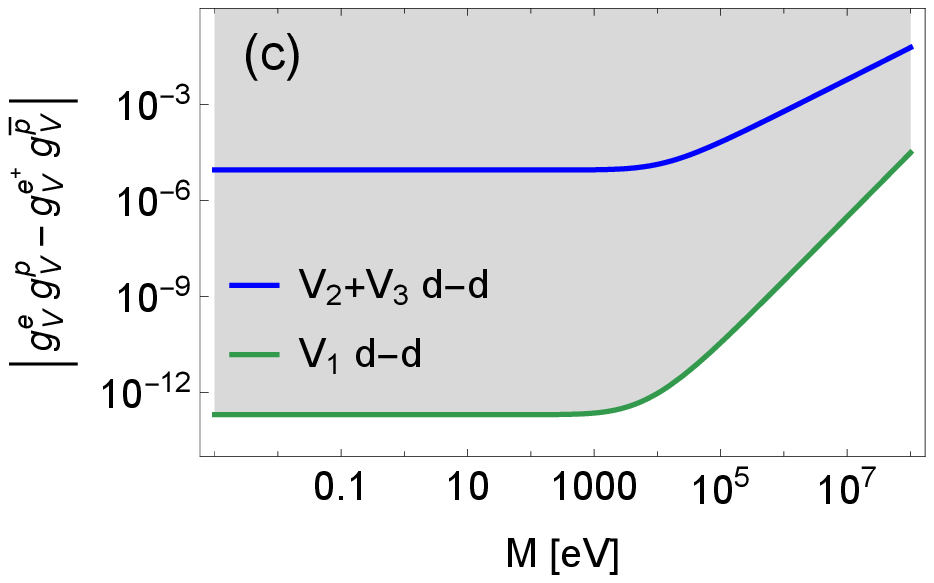}
\caption{\justifying 
Constraints on the coupling constant product differences $g^{e} g^{p} - g^{e^+} g^{\bar{p}}$ for the (a) axial-vector/axial-vector, (b) pseudoscalar/pseudoscalar and (c) vector/vector interactions as a function of the new boson mass $M$ (at 95\% confidence level). 
The $V_1$ d-d results for $g_V g_V$ and $g_s g_s$ have the same value, see Eqs.\,\eqref{gVgV_V1} and \eqref{gsgs_V1}. 
}
\label{fig:CPT}
\end{figure*}

Various methods have been implemented to test the CPT symmetry. 
Current research frequently focuses on direct comparisons between particles and their corresponding antiparticles, examining properties such as mass, charge-to-mass ratio, magnetic moment, and other intrinsic characteristics. Examples include detailed comparisons of protons and antiprotons \cite{gabrielse_precision_1999, smorra_parts-per-billion_2017, borchert_16-parts-per-trillion_2022, ulmer_high-precision_2015}, electrons and positrons \cite{moskal_testing_2021}, as well as hydrogen and antihydrogen \cite{kostelecky_lorentz_2015,charlton_antihydrogen_2020}.
There are additional methods to test CPT violation: atomic clocks \cite{kostelecky_lorentz_2018}; particle-antiparticle oscillation studies which can detect subtle CPT-violating effects through careful analysis of oscillation frequencies and decay rate differences \cite{roberts_testing_2017}; and observations of cosmic rays, photons, and neutrinos over cosmological distances \cite{li_cosmological_2008, alan_kostelecky_lorentz_2004, barenboim_neutrinos_2002}.
Lastly, since CPT symmetry implies Lorentz invariance in quantum field theories, testing for Lorentz invariance violation may also serve as an indirect probe 
of CPT violation
\cite{greenberg_cpt_2002, kostelecky_data_2011}. 
Theories with Lorentz and/or CPT violation have been considered, e.g., in Refs.~\cite{greenberg_cpt_2002,dolgov_cpt_2012,tureanu_cpt_2013,kostelecky_lorentz_2015}.


The effects of exotic interactions on the anti-hydrogen $d-d$ transition, when compared with that in hydrogen, can be used to test the CPT symmetry. 
We take advantage of the fact that, under CPT, the energy intervals in hydrogen and antihydrogen are the same 
in the standard-model based on the principles of quantum field theory,
so a test of CPT invariance can be done by a direct comparison of experimental results for hydrogen and antihydrogen without any reliance on theoretical calculations. 

Considering the $1\textrm{S}_d$ and $2\textrm{S}_d$ wave functions are $\ket {\Psi_{1\textrm{S}_d}}$ and $\ket {\Psi_{2\textrm{S}_d}}$, 
the exotic interaction potentials $V_i$ give a shift of the $d-d$ energy interval $E_{d-d}$: 
\begin{equation}
\begin{aligned}\label{DeltaE}
\Delta E^\textrm{Exotic}_{d-d}=\bra {\psi_{2\textrm{S}_d}} V_i \ket {\psi_{2\textrm{S}_d}}  - \bra {\psi_{1\textrm{S}_d}} V_i \ket {\psi_{1\textrm{S}_d}} \, , 
\end{aligned}
\end{equation}
where the exotic potentials $V_i$ are shown in Eqs.\,\eqref{gaga_V2}-\eqref{gsgs_V1}.
The difference in the energy shift, 
$(\Delta E^\textrm{Exotic}_{d-d})^{\textrm{H}} - (\Delta E^\textrm{Exotic}_{d-d})^{\bar{\textrm{H}}}$,
due to exotic interactions in H and $\bar{\textrm{H}}$
is constrained by the experimental difference,  
${(E_{d-d})}^{\textrm{H}}_{\textrm{exp}} - {(E_{d-d})}^{\bar{\textrm{H}}}_{\textrm{exp}}$. 
The maximal discrepancy between the experimental results of H and $\overline{\textrm{H}}$ at the 95\% confidence level is presented in the right panel of Table\,\ref{table1} as $\Delta E$. 

We therefore obtain constraints on the difference of exotic interaction parameters, $g_i^{e}g_i^{p} - g_i^{e^+}g_i^{\overline p}$.
As shown in Fig.\,\ref{fig:CPT}\,(a-c), the constraints derived from the d-d transition are strong for the difference in $g_A g_A$ between hydrogen and antihydrogen. Specifically, at a new boson mass of 1\,eV, $g_A^{e}g_A^{p} - g_A^{e^+}g_A^{\overline p}$ is restricted to be smaller than $10^{-20}$. 
For the difference in $g_p g_p$, we find a constraint of less than $2\times10^{-5}$ for new boson masses $M<10^3$\,eV. Meanwhile, the constraints arising from the spin-independent term $V_1$ are tight for the respective differences in $g_V g_V$ and $g_s g_s$, which are below $10^{-12}$ for $M<10^3$\,eV.
Importantly, we observe that the constraints on the differences $g_i^{e}g_i^{p} - g_i^{e^+}g_i^{\overline p}$ from hydrogen-antihydrogen comparisons are more stringent than the corresponding limits on $g_i^{e^+}g_i^{\overline p}$ from antihydrogen spectroscopy alone. This is mainly due to the better \textit{absolute} experimental precision of the $d$-$d$ measurement of Ref.~\cite{ahmadi_characterization_2018} in antihydrogen compared with the hyperfine structure interval measurements of Refs.~\cite{ahmadi_observation_2017-1,baker_precision_2025} in antihydrogen.



In summary, we present the study of spin-dependent exotic interactions mediated by spin-0 or spin-1 bosons in antimatter, placing constraints on semileptonic spin-dependent interactions involving positron-antiproton pairs in antihydrogen. To our knowledge, no previous work has explored these interactions in antimatter systems.
We also introduce a novel matter-antimatter comparison test, extending previous CPT violation studies. This involves a direct comparison of the experimental results for hydrogen and antihydrogen.
Since the theoretical accuracy of hyperfine structure calculations significantly exceeds current experimental precision, future advances in antihydrogen spectroscopy, such as ASACUSA-CUSP \cite{malbrunot_asacusa_2018,kraxberger_upgrade_2023,nowak_cpt_2024}, could improve these constraints by orders of magnitude once the experimental precision reaches theoretical precision.

\section*{Acknowledgements}
The authors acknowledge helpful discussion with Jeffrey Hangst, Francis Robicheaux, Stefan Eriksson, and Eric Hunter.
This research was supported in part by the DFG Project ID 390831469: EXC 2118 (PRISMA+ Cluster of Excellence) and by the COST Action within the project COSMIC WISPers (Grant No. CA21106).
The work of Y.\,V.\,S.~was supported by the Australian Research Council under the Discovery Early Career Researcher Award No.~DE210101593. F.\,F.~acknowledges the support of the Austrian Science Fund (FWF) via Project No.~P 36455 (DOI:10.55776/P36455) and Wittgenstein Award (DOI:10.55776/Z387)


\providecommand{\noopsort}[1]{}
%


\appendix
\section{
Appendix A: 
Hyperfine energy level structure of (anti)hydrogen
} 
\label{App1levelstructure}
 
\renewcommand{\theequation}{A\arabic{equation}}
\setcounter{equation}{0}
\renewcommand{\thefigure}{A\arabic{figure}}
\setcounter{figure}{0}

\begin{figure}[!htbp]
\includegraphics[width=0.45\textwidth]{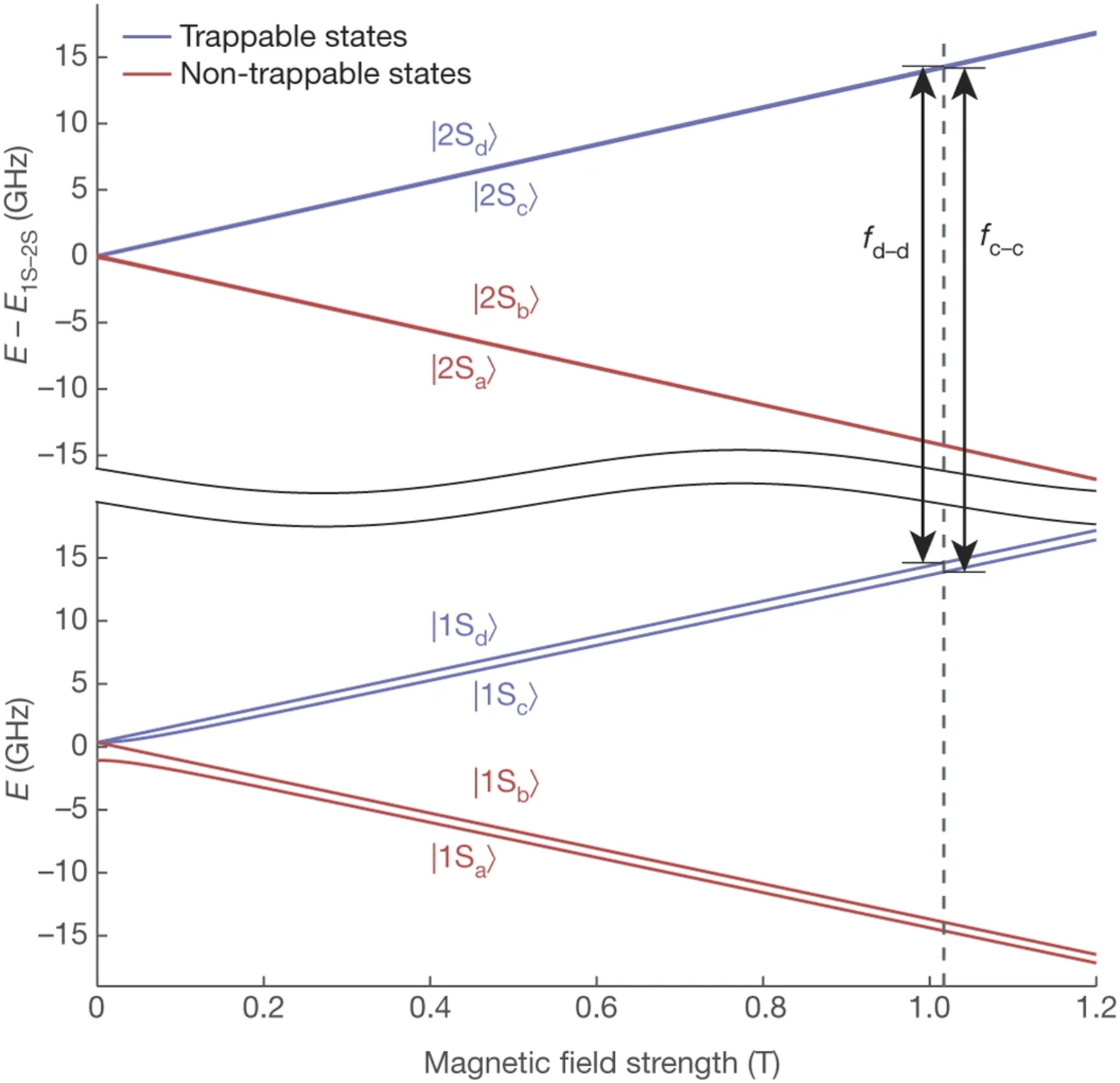}
\caption{\justifying Hydrogenic hyperfine energy levels of the 1S and 2S states in a magnetic field, taken from Ref.\,\cite{ahmadi_observation_2017}. 
The energy differences $E_{\textrm{1S-2S}}$ in the upper section are defined with respect to the centroid energy difference.
}
\label{fig:level-structure}
\end{figure}

The $d$-$d$ transition refers to the transition between $1$S and $2$S hyperfine sublevels, which are consistently labeled as $a, b, c, d$. At zero magnetic field, these states are well-defined by their total angular momentum ($F$) and its projection ($m_F$): The singlet state ($F = 0$) is denoted as $\chi_a$. 
The triplet states ($F = 1$, $m_F=-1, 0, +1$) are denoted as $\chi_{b,\,c,\,d}$. 
In the presence of an external magnetic field, these triplet states split further due to the Zeeman effect \cite{budker_atomic_2010}. 

\section{Appendix B: Constraints on exotic interactions}
\label{appendix_B}

\renewcommand{\theequation}{B\arabic{equation}}
\setcounter{equation}{0}


For an exotic potential $V_i$, to calculate the change of the energy interval between two hyperfine states, we first 
construct the corresponding wavefunctions. 
Taking the 1S hfs interval as an example, the total wavefunction $\Psi_{1\textrm{S}_{a,b,c,d}}(r)$ is given in Eq.\,\eqref{wavefunctionmain}. 
We then calculate the expectation value of an exotic potential $V_i$ to determine the energy shift in each hyperfine state due to this interaction (at zero magnetic field): 
\begin{equation}
\langle V_i \rangle_F = \int \Psi_{F, m_F}^*(r) V_i \Psi_{F, m_F}(r) \, d^3r\,,
\end{equation}
which can be simplified as:
\begin{equation}
\langle V_i \rangle_F = \int \left| \psi_{1\textrm{S}}(r) \right|^2 V_i(r) \, d^3r \, \cdot \, \langle \chi_{F, m_F} | V_i^{\text{spin}} | \chi_{F, m_F} \rangle
\end{equation}
where \( V_i^{\text{spin}} \) represents the spin-dependent part of $V_i$, typically involving the dot product of the electron and nuclear spins. 
For instance, in the case of  $V_2$, \( V_i^{\text{spin}} \) contains terms like \( \boldsymbol{\sigma}_{e^+} \cdot \boldsymbol{\sigma}_{\overline p} \) 
[see Eq.\,\eqref{gaga_V2}], 
the expectation values of
which differ between the \( F = 1 \) and \( F = 0 \) states. 
\( V_i(r) \) is the remaining spin-irrelevant part of \( V_i \). 

The expectation values of \( V_i^{\text{spin}} \) for $V_2$ are: 
\begin{itemize}
\item Singlet state \( F = 0 \): \( \langle \boldsymbol{\sigma}_{e^+} \cdot \boldsymbol{\sigma}_{\overline p} \rangle_{F=0}  = -3 \)\,;
\item Triplet state \( F = 1 \): \( \langle \boldsymbol{\sigma}_{e^+} \cdot \boldsymbol{\sigma}_{\overline p} \rangle_{F=1} = 1 \).
\end{itemize}




As for \(V_3\), it also contains a term involving the projection of each spin onto the radial direction \({\boldsymbol{r}}\), i.e., \(\left( \boldsymbol{\sigma}_{e^+} \cdot \hat{\boldsymbol{r}} \right) \left( \boldsymbol{\sigma }_{\overline p}  \cdot \hat{\boldsymbol{r}} \right) \). 
We can use the Wigner-Eckart theorem and symmetry considerations to evaluate the relevant expectation values. For $s$-wave atomic states (which are spherically symmetric), the expectation values are: 
\begin{itemize}
\item For the singlet state with $F=0$: 
\[
\langle \overline{ \left( \boldsymbol{\sigma}_{e^+} \cdot \hat{\boldsymbol{r}} \right) \left( \boldsymbol{\sigma}_{\overline{p}}  \cdot \hat{\boldsymbol{r}} \right) } \rangle_{F=0} = 
\frac{1}{3} \langle \boldsymbol{\sigma}_{e^+} \cdot \boldsymbol{\sigma}_{\overline{p}} \rangle_{F=0} = -1
\,; 
\]
\item For the triplet state with $F=1$: 
\[
\langle \overline{ \left( \boldsymbol{\sigma}_{e^+} \cdot \hat{\boldsymbol{r}} \right) \left( \boldsymbol{\sigma}_{\overline{p}}  \cdot \hat{\boldsymbol{r}} \right) } \rangle_{F=1} = \frac{1}{3} \langle \boldsymbol{\sigma}_{e^+} \cdot \boldsymbol{\sigma}_{\overline{p}} \rangle_{F=1} =\frac{1}{3} \, , 
\]
\end{itemize}
where in the first equality of both equations, we have taken the angular average over all directions, denoted by the overline.

Then, the energy difference \( \Delta E \) between the hyperfine states due to the exotic potential \( V_i(r) \) is given by:
\begin{equation}
\Delta E^\textrm{Exotic} = \langle V_i \rangle_{F=1} - \langle V_i \rangle_{F=0} \, . 
\end{equation}
$\Delta E^\textrm{Exotic}$ takes the form of a product of a boson-mass-dependent parameter $C(M)$, and the coupling constant product $g^{e^+} g^{\bar{p}}$. 
It is constrained by the maximum
possible deviation $\Delta E$ given in Table\,\ref{table1}. Therefore, for a certain new boson mass M,
\begin{equation}
\left|g^{e^+} g^{\bar{p}}\right| \leq \frac{|\Delta E|}{|C(M)|}. 
\end{equation}

\end{document}